# A hexahedron element formulation with a new multi-resolution analysis

Xia YiMing    Chen ShaoLin

(1.Civil Engineering Department, Nanjing University of Aeronautics and Astronautics, Nanjing 210016, China )

Email: xym4603@sina.com

A multi-resolution hexahedron element and method is presented with a new multi-resolution analysis (MRA) framework. The MRA framework is formulated out of a mutually nesting displacement subspace sequence, whose basis functions are constructed of scaling and shifting on the element domain of a basic node shape function. The basic node shape function is constructed from extending to other seven quadrants around a specific node of a node isoparametric shape function. The MRA endows the proposed element with the resolution level (RL) to adjust the element node number, thus modulating structural analysis accuracy accordingly. As a result, the traditional 8-node hexahedron element and method is a mono-resolution one and also a special case of the proposed element and method. The accuracy of a structural analysis is actually determined by the RL, not by the mesh. The simplicity and clarity of shape function construction with the Kronecker delta property and the rational MRA enable the proposed element method to be more rational, easier and efficient in its implementation than the conventional mono-resolution solid element method or other MRA methods. The multi-resolution hexahedron element is more adapted to dealing with the accurate computation of structural problems.

**Keywords: hexahedron element, multi-resolution analysis (MRA), resolution level (RL), basic node shape function, mutually nesting displacement subspace sequence, scaling and shifting**.
**PACS number(s):** 02.60.Cb, 02.60.Jh, 02.60.Lj, 02.70.Dh

## 1. Introduction

Multi-resolution analysis (MRA) is a widely popular technique that has been applied in many domains such as the signal and image processing, the damage detection and health monitoring, and the differential equation solution. However, in the field of computational mechanics, the MRA has not been, in a real sense, fully utilized in the numerical solution of engineering problems either by the traditional finite element method (FEM) or by other methods such as the wavelet finite element method (WFEM) [2,3], the meshfree method (MFM) [4,5] and the natural element method (NEM) [6,7] etc.

As is commonly known of the FEM, owing to invariance of the node number that a single finite element contains, the finite element can be regarded as a mono-resolution one from a MRA point of view and the FEM structural analysis is usually not associated with the MRA concept. However, it is, in fact, by means of finite element model meshing and re-meshing to modulate analysis accuracy in which a cluster of mono-resolution finite elements are assembled together artificially that the rough structural MRA is executed by the FEM. As a result, in the overall structural analysis process, the traditional finite element model has to be re-meshed until sufficient accuracy is reached, which leads to the low computation efficiency or convergent rate. Since the MRA out of the FEM is irrational and the mono-resolution finite elements are assembled together artificially, the deficiency of the FEM becomes explicit in the accurate computation of structural problems with local steep gradient such as material nonlinearity [8,9], local damage and crack [10,11], impacting and exploding problems [12,13].

The great efforts have been made over the past thirty years to overcome the drawback of the FEM with many improved methods to come up, such as WFEM, MFM and NEM etc., which open up a transition from the monoresolution finite element method to the multiresolution finite element method featured with the adjustable element node number. Although these MRA methods have illustrated their powerful capability and computational efficiency in dealing with some problems, they always have such major inherent deficiencies as the complexity of a node shape function construction, the absence of the Kronecker delta property of the node shape function and the lack of a solid mathematical basis for the MRA, which make the treatment of element boundary condition complicated and the selection of element node layout empirical, that substantially reduce computational efficiency. Thus, these MRA methods have never found a wide application in engineering practice just as the FEM. In fact, they can be viewed as the intermediate products in the transition of the traditional FEM from the monoresolution to the multiresolution.

The drawbacks of all those MRA methods can be removed by the introduction of a new multiresolution finite method in this paper. With respect to a solid element in the finite element stock, a multiresolution hexahedron element is formulated out of a displacement subspace sequence which is constituted by the translated and scaled version as subspace basis functions of the basic node shape function. The basic node shape function is constructed by extending the node isoparametric shape functions of the traditional 8-node hexahedron element to other seven quadrants around a specific node. The various node shape functions within the multiresolution element are easily made of the scaled and shifted version of the basic node shape function. It can be seen that the shape function construction is simple and clear and the node shape functions hold the Kronecker delta property. In addition, the proposed element method possesses a solid mathematical basis for the MRA, which endows the proposed element with the resolution level (RL) that can be modulated to change the element node number, adjusting structural analysis accuracy accordingly. As a result, the proposed multiresolution finite element method can bring about substantial improvement of the computational efficiency in the structural analysis when compared with the corresponding FEM or other MRA methods.

## 2. Basic node shape function construction

An 8-node hexahedron element is very popular in engineering practice due to its versatility and balance of computational accuracy and cost. The following is the crucial step of constructing an 8-node basic node shape function in the formulation of a multi-resolution hexahedron element.

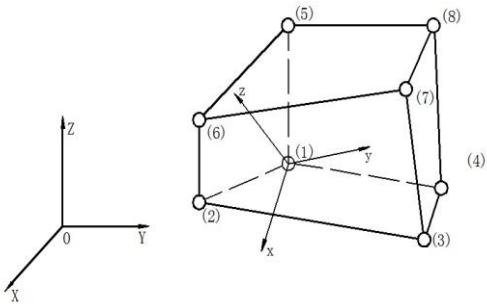 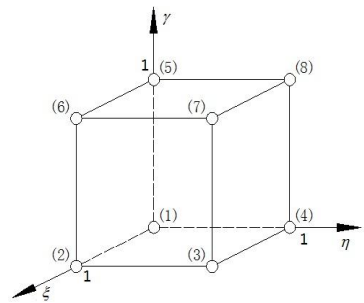

**Fig. 1** An 8-node hexahedron element    **Fig. 2** The space domain for an 8-node isoparametric hexahedron element

Corresponding to an 8-node hexahedron element of arbitrary shape in a Carstesian coordinate system shown in Fig.1, an isoparametric 8-node hexahedron element in a natural coordinate system shown in Fig.2 is adopted with the transforming relations from the Carstesian coordinate system to the natural one($\xi\ \eta\ \gamma$) defined as follows:

$$x = \sum_{i=1}^{8} N_i x_i \quad y = \sum_{i=1}^{8} N_i y_i \quad z = \sum_{i=1}^{8} N_i z_i \tag{1}$$

in which $x_i$, $y_i$, $z_i$ are the Carstesian coordinate values at the $i$-th node.

Thus, the displacements can be easily defined as follows:

$$u^e = \sum_{i=1}^{8} N_i u_i^e \quad v^e = \sum_{i=1}^{8} N_i v_i^e \quad w^e = \sum_{i=1}^{8} N_i w_i^e \tag{2}$$

where $u^e, v^e, w^e$ are the displacements in the $x$, $y$, $z$ directions at an arbitrary point of the element respectively. $u_i^e$, $v_i^e$, $w_i^e$ are the displacements at the $i$-th node respectively. $N_i$ is the conventional shape function for the different node, which are defined on the space domain of $[0,1]^3$ with the tri-linear functions as follows:

$$\begin{cases} N_1(\xi,\eta) = (1-\xi)(1-\eta)(1-\gamma) \\ N_2(\xi,\eta) = \xi(1-\eta)(1-\gamma) \\ N_3(\xi,\eta) = \xi\eta(1-\gamma) \\ N_4(\xi,\eta) = (1-\xi)\eta(1-\gamma) \\ N_5(\xi,\eta) = (1-\xi)(1-\eta)\gamma \\ N_6(\xi,\eta) = \xi(1-\eta)\gamma \\ N_7(\xi,\eta) = \xi\eta\gamma \\ N_8(\xi,\eta) = (1-\xi)\eta\gamma \end{cases} \tag{3}$$

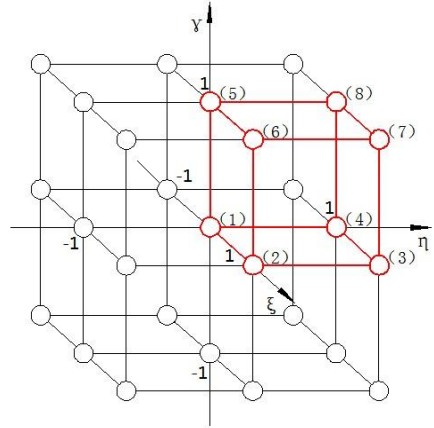

**Fig. 3** The extended space domain for the 8-node isoparametric hexahedron element

For the formulation of a multi-resolution 8-node hexahedron element, as shown in Fig 2., an arbitrary node(1) isoparametric shape function defined on the space domain of $[0, 1]^3$ should be extended to the domain of $[-1,1]^3$ by means of shifting the isoparametric element around the node(1) (at the point (0,0,0)) vertically, horizontally and obliquely to the other seven quadrants, thus covering all 26 nodes adjacent to the node (1) shown in Fig 3, and finally the basic isoparametric extended shape function for node (1) at the coordinate of (0, 0, 0) can be defined as follows:

$$\varphi(\xi,\eta,\gamma) := \begin{cases} N_1(\xi,\eta,\gamma) & \xi \in [0,1], \eta \in [0,1], \gamma \in [0,1] \\ N_2(1+\xi,\eta,\gamma) & \xi \in [-1,0], \eta \in [0,1], \gamma \in [0,1] \\ N_3(1+\xi,1+\eta,\gamma) & \xi \in [-1,0], \eta \in [-1,0], \gamma \in [0,1] \\ N_4(\xi,1+\eta,\gamma) & \xi \in [0,1], \eta \in [-1,0], \gamma \in [0,1] \\ N_5(\xi,\eta,1+\gamma) & \xi \in [0,1], \eta \in [0,1], \gamma \in [-1,0] \\ N_6(1+\xi,\eta,1+\gamma) & \xi \in [-1,0], \eta \in [0,1], \gamma \in [-1,0] \\ N_7(1+\xi,1+\eta,1+\gamma) & \xi \in [-1,0], \eta \in [-1,0], \gamma \in [-1,0] \\ N_8(\xi,1+\eta,1+\gamma) & \xi \in [0,1], \eta \in [-1,0], \gamma \in [-1,0] \\ 0 & \xi \notin [-1,1], \eta \notin [-1,1], \gamma \notin [-1,1] \end{cases} \quad (4)$$

The Kronecker delta property holds for the basic node shape function $\varphi(\xi,\eta,\gamma)$ as follows:

$$\varphi(0,0,0) = 1 \quad \varphi(26 nodes) = 0 \qquad (5)$$

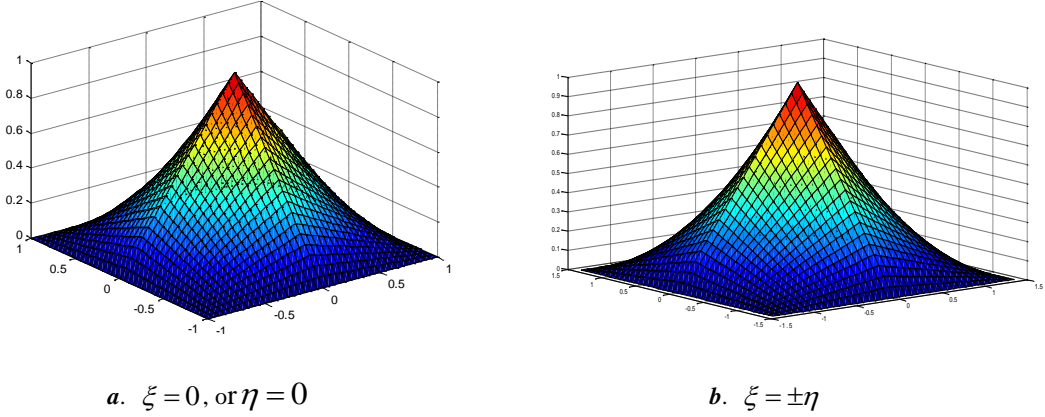

*a*. $\xi = 0$, or $\eta = 0$          *b*. $\xi = \pm\eta$

**Fig .4** The basic node shape function $\varphi(\xi,\eta,\gamma)$ covering 27 nodes on the space domain $[-1, 1]^3$

The basic shape function $\varphi(\xi,\eta,\gamma)$ is shown in Figs. 4 *a, b*. Obviously, the basic node shape function $\varphi(\xi,\eta,\gamma)$ is continuous.

## 3. Basis function construction for a displacement subspace sequence

## forming a new MRA

In order to carry out a MRA of a structure, a mutual nesting displacement subspace sequence for a solid element should be established. In this paper, a totally new technique is proposed to construct the MRA which is based on the concept that a subspace sequence (multi-resolution subspaces) can be formulated by subspace basis function vectors at different resolution levels whose elements- basis function vector can be constructed by scaling and shifting on the space domain of the basic node shape function without using wavelet basis functions. The displacement subspace basis function vector at an arbitrary RL of $(m+1)\times(n+1)\times(l+1)$ (total node number) for a hexahedron element is formulated as follows:

$$\Psi_{mnl} = \begin{bmatrix} \Phi_{mnl,000} \cdots \Phi_{mnl,rst} \cdots \Phi_{mnl,mnl} \end{bmatrix} \tag{6}$$

where $\Phi_{mnl,rst} = \begin{bmatrix} \varphi_{mnl,rst} & 0 & 0 \\ 0 & \varphi_{mnl,rst} & 0 \\ 0 & 0 & \varphi_{mnl,rst} \end{bmatrix}$ is the basis function vector,

$\varphi_{mnl,rst} = \varphi(m\xi - r, n\eta - s, l\gamma - t)$, $m$, $n$, $l$ denoted as the positive integers, the scaling parameters in $\xi$, $\eta$, $\gamma$ directions respectively. $r = 0,1,2,3\cdots m$, $s = 0,1,2,3\cdots n$, $t = 0,1,2,3\cdots l$. $(r,s,t)$ is the shifted node position indicator, $(m\xi - r) \in [-1,1], \xi \in [0,1], (n\eta - s) \in [-1,1], \eta \in [0,1], (l\gamma - t) \in [-1,1], \gamma \in [0,1]$. Here, the space domain of the variables is retrieved to the initial $[0,1]^3$.

It is seen from Eq. (6) that the nodes for the scaling process are equally spaced on the space domain $[0,1]^3$ with a step size of $1/m$ in $\xi$, $1/n$ in $\eta$, $1/l$ in $\gamma$ directions respectively.

Scaling of the basic isoparametric extended shape function on the space domain of $[-1,1]^3$ (precisely on the domain of $[\frac{-1}{m},\frac{1}{m}]\times[\frac{-1}{n},\frac{1}{n}]\times[\frac{-1}{l},\frac{1}{l}]$) and then shifting to other nodes $\left(\frac{r}{m},\frac{s}{n},\frac{t}{l}\right)$ on the domain of $[0,1]^3$ respectively will produce the various shifted node shape functions, which are shown in Fig5. at the RL of $2\times2\times2$, $3\times3\times3$.

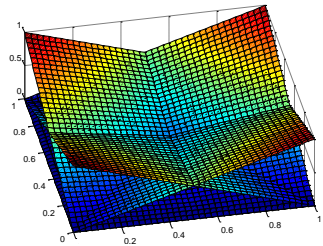
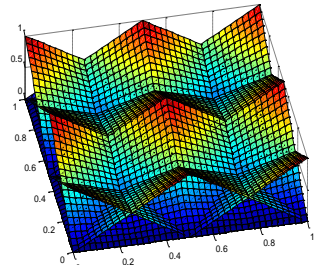

*a.* $\varphi(\xi,\eta,\gamma) \cdot \xi = 0$, or $\eta = 0$ (RL=2×2)  　　　　*b.* $\varphi(\xi,\eta,\gamma) \cdot \xi = 0$, or $\eta = 0$ (RL=3×3)

**Fig. 5** The scaled and shifted version of the basic node shape function $\varphi(\xi,\eta,\gamma)$ on the element domain $[0,1]\times[0,1]\times[0,1]$

Since the elements of basis function vector are linearly independent respectively of the various scaling and the different shifting parameters, the subspaces in the subspace sequence established are mutually nested, thus formulating a MRA framework, that is.

$$\begin{aligned}
\mathbf{W}_{mnl} &= \begin{bmatrix} V_{111} & \cdots & V_{ijk} & \cdots & V_{mnl} \end{bmatrix} \\
V_{ijk} &:= span\{\mathbf{\Psi}_{ijk} : i,j,k \in Z\} \\
V_{ijk} &\subset V_{(i+1)jk} \quad V_{ijk} \subset V_{i(j+1)k} \quad V_{ijk} \subset V_{ij(k+1)} \\
V_{ijk} &\subset V_{(i+1)(j+1)k} \quad V_{ijk} \subset V_{(i+1)j(k+1)} \quad V_{ijk} \subset V_{(i+1)(j+1)(k+1)}
\end{aligned} \tag{7}$$

where $Z$ is denoted as the positive integers, $V_{ijk}$ as a displacement subspace at the *RL* of $(i+1)\times(j+1)\times(k+1)$.

Thus, it can be seen that the mutually nesting displacement subspace sequence $\mathbf{W}_{mnl}$ can be taken as a solid mathematical foundation for the MRA framework that makes the new MRA rational and $V_{111}$ is equivalent to the displacement field for a traditional solid element, That is why the traditional hexahedron element is regarded as a mono-resolution one and also a special case of the multi-resolution solid element.

Based on the newly established MRA, the displacements in *x, y, z* directions at an arbitrary point of an 8-node hexahedron element in the displacement subspace at the RL of $(m+1)\times(n+1)\times(l+1)$ can be defined as follows

$$\mathbf{\Delta}_{mnl} = \begin{Bmatrix} u^e_{mnl} \\ v^e_{mnl} \\ w^e_{mnl} \end{Bmatrix} = \mathbf{\Psi}_{mnl} \mathbf{a}^e_{mnl} \tag{8}$$

where $\mathbf{a}^e_{mnl} = \begin{bmatrix} [u_{000}, v_{000}, w_{000}] \ldots & [u_{rst}, v_{rst}, w_{rst}] \ldots & [u_{mnl}, v_{mnl}, w_{mnl}] \end{bmatrix}^T$, $u_{rst}, v_{rst}, w_{rst}$ are the displacements in *x, y, z* directions respectively at the element node $\left(\dfrac{r}{m}, \dfrac{s}{n}, \dfrac{t}{l}\right)$

It can be seen that the proposed multi-resolution element is meshfree, whose nodes are uniformly scattered at each coordinate, node number and node position fully determined by the *RL*. When the scaling parameter $m=n=l=1$ ( $RL=2\times2\times2$), that is a traditional 8 node hexahedron element, Eq. (8) will become Eq. (2).

## 4. A hexahedron element formulation with MRA

The generalized function of potential energy in the displacement subspace at the resolution level of $(m+1)\times(n+1)\times(l+1)$ for a hexahedron element is

$$\Pi_p(V_{mnl}) = \frac{1}{2}\int_V [\varepsilon]_{mnl}[D][\varepsilon]_{mnl}^T dV - \frac{1}{2}\rho\omega^2 \int_V \mathbf{\Delta}_{mnl}\mathbf{\Delta}_{mnl}^T dV \qquad (9)$$
$$- \int_V \mathbf{p}\mathbf{\Delta}_{mnl} dV - \int_S \mathbf{q}\mathbf{\Delta}_{mnl}^s dS - \sum_i \mathbf{P}_i \mathbf{\Delta}_{mnl}(\xi_i, \eta_i, \gamma_i)$$

where $[\varepsilon]_{mnl}^e = \left[\dfrac{\partial u_{mnl}^e}{\partial x} \quad \dfrac{\partial v_{mnl}^e}{\partial y} \quad \dfrac{\partial w_{mnl}^e}{\partial z} \quad \dfrac{\partial u_{mnl}^e}{\partial y}+\dfrac{\partial v_{mnl}^e}{\partial x} \quad \dfrac{\partial u_{mnl}^e}{\partial z}+\dfrac{\partial w_{mnl}^e}{\partial x} \quad \dfrac{\partial v_{mnl}^e}{\partial z}+\dfrac{\partial w_{mnl}^e}{\partial y}\right]^T$,

$$[D] = C\begin{bmatrix} 1 & & & & & \\ \frac{\mu}{1-\mu} & 1 & & sym & & \\ \frac{\mu}{1-\mu} & \frac{\mu}{1-\mu} & 1 & & & \\ 0 & 0 & 0 & \frac{1-2\mu}{2(1-\mu)} & & \\ 0 & 0 & 0 & 0 & \frac{1-2\mu}{2(1-\mu)} & \\ 0 & 0 & 0 & 0 & 0 & \frac{1-2\mu}{2(1-\mu)} \end{bmatrix}$$, $C = \dfrac{E(1-\mu)}{(1+\mu)(1-2\mu)}$, $E$ is the

material Young modulus, $\omega$ angular frequency, $\mathbf{p}$ distributed volume force vector, $\mathbf{q}$ distributed force vector applied on a surface, and $\mathbf{P}_i$ lump force vector acting at the element point $i$.

From Eq.(8), we can obtain the following formula

$$[\varepsilon]_{mnl} = [B_{000}, \cdots B_{rst}, \cdots B_{mnl}]\mathbf{a}_{mnl}^e \qquad (10)$$

By substituting Eq.(10) into Eq.(9), the concise expression can be obtained after being reassembled as follows:

$$\Pi_p(V_{mnl}) = \frac{1}{2}\mathbf{a}_{mnl}^{eT}\mathbf{K}_{mnl}^e\mathbf{a}_{mnl}^T - \frac{1}{2}\omega^2 \mathbf{a}_{mnl}^{eT}\mathbf{M}_{mnl}^e\mathbf{a}_{mnl}^e - \mathbf{a}_{mnl}^{eT}\mathbf{f}_{mnl}^e - \mathbf{a}_{mnl}^{eT}\mathbf{F}_{mnl}^e \qquad (11)$$

where $\mathbf{K}_{mnl}^e$, $\mathbf{M}_{mnl}^e$ are the element stiffness, mass matrix respectively, $\omega$ is the circular frequency of the element, $\mathbf{f}_{mnl}^e$ the element distributed loading column vector, and $\mathbf{F}_{mnl}^e$ the element lump loading column vector.

According to the potential energy minimization principle, let $\delta\Pi_p(V_{mnl}) = 0$, and the 8-node solid element equilibrium equations can be obtained as follows:

(1) Basic equation

$$\mathbf{K}_{mnl}^e \mathbf{a}_{mnl}^e = \mathbf{f}_{mnl}^e + \mathbf{F}_{mnl}^e \qquad (12)$$

(2) Basic equation of free vibration

$$\left(\mathbf{K}^e_{mnl} - \omega^2 \mathbf{M}^e_{mnl}\right)\mathbf{a}^e_{mnl} = \mathbf{0} \qquad (13)$$

The element expressions of the stiffness matrix $\mathbf{K}^e_{mnl}$, $\mathbf{M}^e_{mnl}$, and the loading column vectors $\mathbf{f}^e_{mnl}$, $\mathbf{F}^e_{mnl}$ can be given as follows:

$$\mathbf{K}^e_{mnl} = \begin{bmatrix} \mathbf{k}^{000}_{000} & \cdots & \mathbf{k}^{000}_{rst} & \cdots & \mathbf{k}^{000}_{mnl} \\ \vdots & & \vdots & & \vdots \\ \mathbf{k}^{rst}_{000} & \cdots & \mathbf{k}^{rst}_{rst} & \cdots & \mathbf{k}^{rst}_{ijk} \\ \vdots & & \vdots & & \vdots \\ \mathbf{k}^{mnl}_{000} & \cdot & \mathbf{k}^{mnl}_{rst} & \cdot & \mathbf{k}^{mnl}_{mnl} \end{bmatrix} \qquad (14)$$

where the superscript is denoted as the row number of the matrix and the subscript as the aligned element node numbering ($r$, $s$, $t$). In terms of the properties of the extended shape functions, we have

$$\begin{cases} \mathbf{k}^{rst}_{rst} = \displaystyle\sum_{\substack{|c-r|\le 1 \\ |d-s|\le 1 \\ |e-t|\le 1}} \mathbf{k}_{cde,rst} \\ \mathbf{k}^{rst}_{rst} = \mathbf{k}_{cde,rst} = 0, \text{ when } |c-r|>1, |d-s|>1, |e-t|>1 \end{cases} \qquad (15)$$

in which $\mathbf{k}_{cde,rst}$ is the coupled node stiffness relating the node ($c$, $d$, e) to the ($r$, $s$, t).

$$\mathbf{k}_{cde,rst} = \int_0^1 \int_0^1 \int_0^1 [B_{cde}]^T [D][B_{rst}] |\mathbf{J}| d\xi d\eta d\gamma \qquad (16)$$

$$\mathbf{M}^e_{mnl} = \int_0^1 \int_0^1 \int_0^1 \mathbf{\Psi}^T_{mnl} \mathbf{\Psi}_{mnl} \rho |\mathbf{J}| d\xi d\eta d\gamma \qquad (17)$$

where $\mathbf{J}$ is denoted as the Jacobian matrix.

$$\mathbf{J} = \begin{bmatrix} \dfrac{\partial x}{\partial \xi} & \dfrac{\partial y}{\partial \xi} & \dfrac{\partial z}{\partial \xi} \\ \dfrac{\partial x}{\partial \eta} & \dfrac{\partial y}{\partial \eta} & \dfrac{\partial z}{\partial \eta} \\ \dfrac{\partial x}{\partial \gamma} & \dfrac{\partial y}{\partial \gamma} & \dfrac{\partial z}{\partial \gamma} \end{bmatrix} \qquad (18)$$

$$\begin{cases} \mathbf{f}^e_{mnl} = \mathbf{f}^{ep}_{mnl} + \mathbf{f}^{eq}_{mnl} \\ \mathbf{f}^{ep}_{mnl,rst} = \int_0^1 \int_0^1 \int_0^1 \left[ \varphi_{mnt,rst} p_x, \varphi_{mnt,rst} p_y, \varphi_{mnt,rst} p_z \right]^T |\mathbf{J}| d\xi d\eta d\gamma \\ \mathbf{f}^{eq}_{mnl,rst} = \int_0^1 \int_0^1 \left[ \varphi_{mnt,rst} q_x, \varphi_{mnt,rst} q_y, \varphi_{mnt,rst} q_z \right]^T \begin{Bmatrix} R_x \\ R_y \\ R_z \end{Bmatrix} d\xi d\eta \\ \mathbf{F}^e_{mnl,rst} = \sum_i \left[ \varphi(m\xi_i - r, n\eta_i - s, l\gamma_i - t) P_{xi}, \varphi(m\xi_i - r, n\eta_i - s, l\gamma_i - t) P_{yi} \right. \\ \qquad \left. \varphi(m\xi_i - r, n\eta_i - s, l\gamma_i - t) P_{zi} \right]^T \end{cases} \quad (19)$$

where $R_x = \begin{vmatrix} \frac{\partial y}{\partial \xi} & \frac{\partial z}{\partial \xi} \\ \frac{\partial y}{\partial \eta} & \frac{\partial z}{\partial \eta} \end{vmatrix}$, $R_y = \begin{vmatrix} \frac{\partial z}{\partial \xi} & \frac{\partial x}{\partial \xi} \\ \frac{\partial z}{\partial \eta} & \frac{\partial x}{\partial \eta} \end{vmatrix}$, $R_z = \begin{vmatrix} \frac{\partial x}{\partial \xi} & \frac{\partial y}{\partial \xi} \\ \frac{\partial x}{\partial \eta} & \frac{\partial y}{\partial \eta} \end{vmatrix}$, $\xi_i$, $\eta_i$, $\gamma_i$ is the local coordinate at the locations the lump loading acting on.

## 5. Transformation matrix

In order to carry out the structural analysis, the element stiffness and mass matrices $\mathbf{K}^e_{mnl}$, $\mathbf{M}^e_{mnl}$, the loading column vectors $\mathbf{f}^e_{mnl}$, $\mathbf{F}^e_{mnl}$ should be transformed from the element local coordinate system (*xyz*) to the structural global coordinate system (*XYZ*) shown in Fig.1. The transforming relations from the local to the global are defined as follows:

$$\mathbf{K}^i_{mnl} = \mathbf{T}^{eT}_{mnl} \mathbf{K}^e_{mnl} \mathbf{T}^e_{mnl} \quad (20)$$

$$\mathbf{M}^i_{mnl} = \mathbf{T}^{eT}_{mnl} \mathbf{M}^e_{mnl} \mathbf{T}^e_{mnl} \quad (21)$$

$$\mathbf{f}^i_{mnl} = \mathbf{T}^{eT}_{mnl} \mathbf{f}^e_{mnl} \quad (22)$$

$$\mathbf{F}^i_{mnl} = \mathbf{T}^{eT}_{mnl} \mathbf{F}^e_{mnl} \quad (23)$$

where $\mathbf{K}^i_{mnl}$, $\mathbf{M}^i_{mnl}$ are the element stiffness and mass matrices respectively, and $\mathbf{f}^i_{mnl}$, $\mathbf{F}^i_{mnl}$ the element loading column vectors under the global coordinate system. $\mathbf{T}^e_{mnl}$ is the element transformation matrix defined as follows:

$$\mathbf{T}^e_{mnl} = \begin{bmatrix} \boldsymbol{\lambda}_{111} & & & \mathbf{0} \\ & \cdots & & \\ & & \boldsymbol{\lambda}_{ijk} & \\ & & & \cdots \\ \mathbf{0} & & & \boldsymbol{\lambda}_{mnl} \end{bmatrix} \qquad \boldsymbol{\lambda}_{ijk} = \begin{bmatrix} \cos\theta_{xX} & \cos\theta_{xY} & \cos\theta_{xZ} \\ \cos\theta_{yX} & \cos\theta_{yY} & \cos\theta_{yZ} \\ \cos\theta_{zX} & \cos\theta_{zY} & \cos\theta_{zZ} \end{bmatrix}$$

where $\theta$ is the intersection angle between the local and the global coordinate axes.

The structural global stiffness, mass matrices $\mathbf{K}_{mn}$, $\mathbf{M}_{mnl}$, the global loading column vectors $\mathbf{f}_{mnl}$, $\mathbf{F}_{mnl}$ can be obtained by splicing.

## 6. Numerical examples

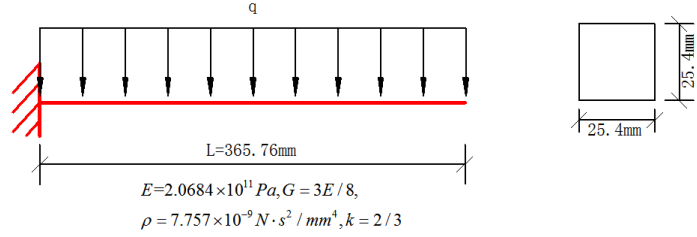

**Fig. 6.** A cantilever beam under the uniform transverse loading $q$.

**Example 1.** A cantilever beam is subjected to the uniform transverse loading $q=1KN/m^2$, with the geometric configuration and the physical parameters shown in **Fig**.6. Evaluate the tip deflection and the basic natural frequency of the beam.

The tip deflection and the basic transverse frequency of this beam are herein solved by the proposed multi-resolution hexahedron element and the conventional 8-node hexahedron element respectively. The RLs of the proposed element and the corresponding meshes of the conventional are compared. The tip deflection and the basic transverse frequency comparisons of the beam are summarized in **table 1**. The RLs or the corresponding meshes are adopted differently. The mesh of $1\times1\times10$ means that mesh of $1\times1$ is for a cross section and 10 is for the longitude of the beam. It can be seen that the computational accuracy with one proposed hexahedron element, which is adjustable through the RL, is equal to that of an arbitrary number of the traditional mono-resolution 8-node hexahedron elements under the same node distribution. Here, the RL adjusting is more easily implemented than the meshing and the re-meshing which means that the proposed element method is more adaptable to the adaptive analysis of the structure. Furthermore, the stiffness, mass matrices and the loading column vectors of one proposed element can be obtained automatically around the nodes and are the same as those of the traditional 8-node hexahedron elements obtained by the artificially complex reassembling around the elements. Thus, the computational efficiency of the proposed element method is higher than the traditional one. The analysis accuracy of one proposed hexahedron element is gradually improved with its RL reaching high and finally the deflection approaches very close to the analytical value. However, the basic beam transverse frequency responses are not convergent to the analytical. The reason behind is that the analytical is based on the 1D Timoshenko beam theory that is less accurate than the 3D element model in the beam dynamic response analysis. In this way, the proposed hexahedron element exhibits its strong capability of accuracy adjustment and its high power of resolution to identify details (nodes) of deformed structure by means of modulating its resolution level, just as a multi-resolution 3D camera with a pixel in its taken photo as a node in the proposed element. Thus, an element of superior analysis accuracy surely has more nodes when compared with that of the inferior, just as a clearer photo contains more pixels.

**Table.1**. The tip defection and the basic transverse frequency of the cantilever beam

| Element type The proposed (RL/one element) | The conventional (mesh) | Tip deflection ($mm$) | Basic frequency ($Hz$) |
|---|---|---|---|
| $2\times 2\times 11$ | $1\times 1\times 10$ | 0.0783 | 1002 |
| $2\times 2\times 21$ | $1\times 1\times 20$ | 0.0788 | 980 |
| $2\times 2\times 31$ | $1\times 1\times 30$ | 0.0790 | 976 |
| analytical value | | 0.0796 [14] | 987 [15] |

**Example 2.** A simply supported square plate with the geometric configuration of length $L$=2m, the thickness $h$=0.2m, the Poisson's ratio $\mu=0.3$ and subjected to the uniform loading $q$=1000N/m². Evaluate the deflections at the center point of the plate.

The displacement responses are found by the proposed hexahedron element, the corresponding traditional 8-node solid element and the wavelet element based on two-dimensional tensor product B-spline wavelet on the interval (BSWI) [2] respectively. The BSWI is chosen because it is the best one among all existing wavelets in approximation of numerical calculation [16] and directly constructed by the tensor product of the wavelets expansions at each coordinate. The central deflections of the plate are summarized in table.2. One proposed multiresolution element with the RLs of $16\times 16\times 2$ and $11\times 11\times 2$ is adopted respectively while one 2D BSWI element of the jth scale=3, the mth order =2, 4 is employed respectively abbreviated as BSWI2$_3$, BSWI4$_3$ with the DOF of $9\times 9$ and $11\times 11$. The RL of each proposed and the corresponding meshes of the conventional are compared. It can be seen that the analysis accuracy with the proposed hexahedron element and the 2D BSWI element are gradually improved with the RL and the order reaching high. It can be seen that the RL adjustment is much easier than the order changing. Although the BSWI4$_3$ is of high accuracy, when compared with the proposed, the deficiency of the BSWI element is obvious. Firstly, in light of tensor product formulation of the multi-dimensional MRA framework, the DOF of a multi-dimensional BSWI element will be so drastically increased from that of a one-dimensional element in an irrational way that the 3D BSWI element is too complex to be applied, resulting in the complex node shape functions and substantial reduction of the computational efficiency. Secondly, the BSWI has no isoparametric element, which restricts the extent of element application.

**Table.2**. Deflection ($w/ qL^4 /100D_0$) at the center point of the simply- supported plate

| Element type The proposed (RL/one element) | The conventional (mesh) | deflection | BSWI[2] | deflection |
|---|---|---|---|---|
| $11\times 11\times 2$ | $10\times 10\times 1$ | 0.4210 | BSWI2$_3$ | 0.3510 |
| $16\times 16\times 2$ | $15\times 15\times 1$ | 0.4257 | BSWI4$_3$ | 0.4273 |
| exact[19] | | 0.4273 | | |

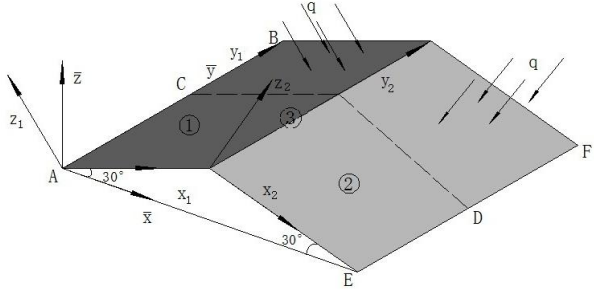 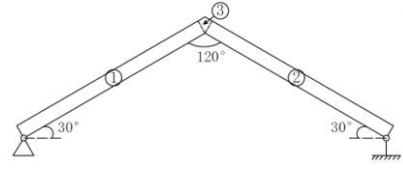

**Fig. 7.** A folded plate under the uniform loading $q$.  **Fig. 8.** The element numbering for a folded plate

**Example 3.** A folded square plate is subjected to the uniform loading $q=1KN/m^2$ shown in **Fig.**7 with its boundary conditions as: boundary lines *AB,EF* are free and the rest are simply supported, and its geometry and physical parameters as: the thickness $h$=1m, length $L$=50m, the elasticity modulus $E= 2.06\times 10^5 Mpa$, the Poisson's ratio $\mu=0.3$. Find the displacement responses of the plate along the boundary line *AB* and midline *CD*.

To calculate the displacement responses requires the folded plate to be herein partitioned into three multi-resolution solid elements ①, ②, ③ with its RL of each element displayed in **Fig.**8. and also into the corresponding meshes with the conventional solid element. In addition, two 2D BSWI4$_3$ elements are also adopted to solve this problem in the literature [3].In the analysis process, these three multi-resolution elements are spliced together along the common intersection boundary and the analysis accuracy can be modulated by means of adjusting the resolution level of the different element. With respect to the conventional, the structure is meshing into a group of mono-resolution elements and these elements are assembled together and the analysis accuracy is improved only by means of re-meshing. It can be seen that the RL adjusting is more rationally and easily implemented than the re-meshing. In addition, the proposed element model of the folded plate structure contains much fewer elements than the traditional element model, thus requiring much less times of the transformation matrix multiplying, which results in much higher computational efficiency for the proposed element method than that for the traditional element method. Moreover, when compared with the proposed element, the 2D BSWI4$_3$ element has the following drawback: due to the absence of Kronecker delta property of the tensor-product constructed shape functions, the special treatments should be taken to deal with the element boundary condition, that is, for the wavelet element, transforming the element deflection field represented by the coefficients of wavelet expansions from wavelet space to physical space, which will bring about low computational efficiency. It would be difficult to find the stable reverse matrix of the transformtion matrix when the dimensional number or the scale becomes higher and the conditional number of the matrix becomes larger. The RL of each proposed and the corresponding meshes of the conventional are compared. The displacement response curves along the boundary line *AB* and midline *CD* are plotted in **Figs 9 and 10**. Case *A* corresponds to RL=16×16×2 of the proposed elements①, ②, that is the mesh of 15×15×1 by the conventional element, Case B corresponds to RL=26×26×2 of the proposed elements①, ②, that is the mesh of 25×25×1 by the conventional. The RL of the intersection boundary should be the same as that of the adjacent element just as P.S the three photos. It can be found that the analysis accuracy with the proposed hexahedron element is gradually improved with its RL approaching high.

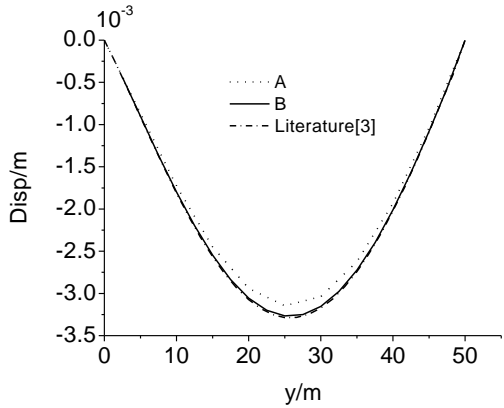
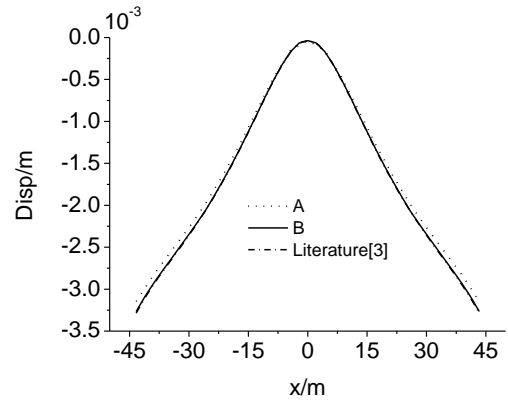

**Fig. 9.** The deflection curve along boundary line AB       **Fig. 10.** The deflection curve along midline CD

## 7. Discussion

The three numerical examples above show that based on the multi-resolution 8-node hexahedron element formulation, the multi-resolution hexahedron element method is introduced, which incorporates such main steps as RL adjusting, element matrix formation, element matrix transformation from a local coordinate system to a global one and global structural matrix formation by splicing of the element matrices. Owing to the existence of the new MRA framework (eq.(8)), the RL adjusting for the proposed element method is more rationally and easily implemented than the meshing and re-meshing for the traditional 8-node hexahedron element method. Due to the extension of the isoparametric node shape function, the stiffness, mass matrices and the loading column vectors of a proposed element can be automatically acquired through quadraturing around nodes in the element matrix formation step while those of the traditional 8-node hexahedron elements are obtained through complex artificially reassembling of the element matrix around the elements in the re-meshing process, which helps improving computation efficiency improvement of the proposed method. Moreover, since the multiresolution hexahedron element model of a structure usually contains much fewer elements than the traditional monoresolution element model, thus requiring much less times of transformation matrix multiplying, the proposed method computes more efficiently than the traditional in the step of element matrix transformation. In addition, because of the simplicity and clarity of isoparametric node shape function formulation with the Kronecker delta property and the rationality of new MRA framework, the proposed method is also superior to other corresponding MRA methods in terms of the computational efficiency, application flexibility and extent. Given all reasons, the multi-resolution hexahedron element method is more rational, easier and efficient in execution, compared with the traditional 8-node hexahedron element method or other corresponding MRA methods, and the proposed solid element is the most accurate one formulated ever since.

## 8. Conclusions

A hexahedron element with a new multi-resolution analysis that has both high power of resolution and strong flexibility of analysis accuracy is introduced into the field of structural analysis. The basic node shape functions are established by shifting to other seven quadrants around a specific node of a basic element in one quadrant and joining the corresponding node shape functions of eight elements at the specific node. A mutually nesting displacement space sequence is constituted from scaling and shifting of the basic node shape functions. As a result, a hexahedron element with a new multi-resolution is derived and the corresponding element method is naturally introduced. The subspace sequence constitutes a new MRA framework and the various node shape functions with the Kronecker delta property coming from the scaled and shifted version of the basic isoparametric node shape functions connect all node-related sub-elements together as a whole. The MRA framework endows the solid element with the RL to adjust the element node number, modulating the analysis accuracy of structure accordingly, while the traditional solid element handles structural node number and analysis accuracy with aid of meshing and re-meshing. It can be seen that the RL adjusting, with its clear mathematical sense, is more rational and easier in implementation than the traditional solid element meshing because there is no MRA framework for the meshing. Thus, the accuracy of a structural analysis is actually determined by the RL, not by the mesh and the adaptive analysis of structure can be handled more conveniently by the proposed method. Furthermore, it is proved that the computational efficiency of the proposed hexahedron element method is much higher than that of the traditional 8-node hexahedron element method or other MRA methods, the conventional solid element and method is a mono-resolution one and can be taken for a special case of the proposed element and method. An element of superior analysis accuracy surely contains more nodes when compared with that of the inferior. Hence, the 8-node hexahedron element with the new MRA is more adapted to dealing with the accurate computation of structural problems with local steep gradient. With advent of the element with the new MRA, the rational MRA will find a wide application in numerical solution to engineering problems in a real sense.

The upcoming work will be focused on the treatment of interface between multiresolution elements of different RL. The interface may be extended to the bridging domain in which the transitional element (expanded Serendipity element) could be used just as P.S photos of different RL.

## 9. Acknowledgements

The authors would like to thank associate editor Prof Li Shaofan at the University of California-Berkeley and two anonymous referees for their valuable comments for improving the quality and the readability of the initial manuscript. Xia is supported by the foundation of Municipal Key Laboratory of Geomechanics and Geological Environment Protection at Chongqing Institute of Logistics Engineering ( Grant No GKLGGP 2013–02). Chen is supported by the National Nature Science Foundation of China ( Grant No 51178222 )